\documentclass[conference]{IEEEtran}

\usepackage{tikz}
\usetikzlibrary{arrows,shapes}
\tikzset{
treenode/.style = {align=center, inner sep=0pt, text centered,
font=\sffamily},
}
\usepackage{epsfig}
\usepackage{pifont}
\usepackage{color}
\usepackage{graphicx}
\usepackage{bm}
\DeclareGraphicsExtensions{.png}
\usepackage{stfloats}
\usepackage{amsfonts}
\usepackage{amsbsy}
\usepackage{amsmath, amsthm, amssymb}
\usepackage{latexsym}
\usepackage{url}
\usepackage{cite}
\usepackage{color}
\usepackage{framed}
\usepackage{algorithm}
\usepackage{algpseudocode}
\usepackage{caption}
\usepackage{subcaption}
\usepackage{soul}
\usepackage{verbatim}

\DeclareCaptionFormat{myformat}{#3}
\newtheorem{assumption}{Assumption}

\newcommand{\Gcal}{{\cal G}}
\newcommand{\Hcal}{{\cal H}}

\newcommand{\Mcal}{{\cal M}}

\newcommand{\Pcal}{{\cal P}}

\newcommand{\Scal}{{\cal S}}

\newcommand{\lhat}{\hat{l}}

\newcommand\blfootnote[1]{%
  \begingroup
  \renewcommand\thefootnote{}\footnote{#1}%
  \addtocounter{footnote}{-1}%
  \endgroup
}

\begin{document}

\title{Optimal Sensor Placement for Topology Identification in Smart Power Grids \\
\normalsize Invited Presentation}

\author{\IEEEauthorblockN{Ananth Narayan Samudrala}
\IEEEauthorblockA{\textit{Lehigh University}\\
ans416@lehigh.edu}
\and
\IEEEauthorblockN{M. Hadi Amini}
\IEEEauthorblockA{\textit{Carnegie Mellon University}\\
mamini1@andrew.cmu.edu}
\and
\IEEEauthorblockN{Soummya Kar}
\IEEEauthorblockA{\textit{Carnegie Mellon University}\\
soummyak@andrew.cmu.edu}
\and
\IEEEauthorblockN{Rick S. Blum}
\IEEEauthorblockA{\textit{Lehigh University}\\
rblum@lehigh.edu}
}

\maketitle

\begin{abstract}
Accurate network topology information is critical for secure operation of smart power distribution systems. Line outages can change the operational topology of a distribution network. As a result, topology identification by detecting outages is an important task to avoid mismatch between the {topology that the operator believes is present and the actual topology}. Power distribution systems are operated as radial trees and are recently adopting the integration of sensors to monitor the network in real time. In this paper, an optimal sensor placement solution is proposed that enables outage detection through statistical tests based on sensor measurements. Using two types of sensors, node sensors and line sensors, we propose a novel formulation for the optimal sensor placement as a  cost optimization problem with binary decision variables, i.e.,  {to place or not place a sensor at each bus/line}. The advantage of the proposed placement strategy for outage detection is that it incorporates various types of sensors,  is independent of load forecast statistics and is cost effective. Numerical results illustrating the placement solution are presented. 

\end{abstract}


%

\section{Introduction}
\blfootnote{This work was supported by the Department of Energy under Award DE-OE0000779.}
Real-time knowledge of the network topology is critical in distribution networks since certainty about the actual topology enables situational awareness which is important for preventing customer outages, and control and dispatch operations\cite{power_book}. Knowledge of the current operational topology of the distribution network is a key building block for various grid monitoring, control, and protection tasks, such as optimal operation of distributed energy resources and microgrids\cite{monitoring_DER,sectionalization, demand_response}. There is a significant dearth of systematic methods for holistic real-time monitoring of the distribution network topology in the face of adversarial cyber-attacks; new technical and domain-specific challenges have to be addressed in detecting and protecting the network against such attacks. Efficient data-integrative modeling is required to achieve optimal placement of micro phasor measurement units ($\mu$PMUs) and power flow sensors to enable topological observability and reliable topology state monitoring with meaningful performance guarantees.

\begin{figure} [h]
\begin{center}
\includegraphics[width = 0.5\textwidth]{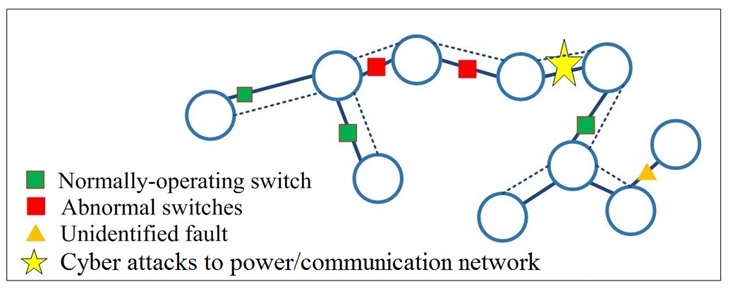}
\caption{Causes for a distribution network topology change}\label{gereric}
\end{center}
\end{figure}

Distribution networks mostly have radial topology and one-way power flows, leading to limited usage and traditionally limited deployment of measurement devices beyond the substation. However in recent years, increasing load and consumption demands is pushing the use of advanced measurement devices to monitor the grid. Several types of line sensors are being developed by various manufacturers, each with unique monitoring capabilities. Also, high precision phasor measurement units ($\mu$PMUs) are being developed to provide synchronized data of high accuracy\cite{von}. In the literature, using measurement data from these advanced sensors several studies have been developed for important distribution network tasks such as distribution system state estimation (DSSE) \cite{DSSE_lu,DSSE_singh,DSSE_baran}. However, all these studies assume that the operator has perfect knowledge of the current topology of the network. But topology attacks or failures in the distribution network change the operational topology of the distribution network, and this affects the network operator's ability to perform important tasks. 
Figure \ref{gereric} illustrates various instances that can lead to a change in the topology of the distribution network such as unidentified faults, cyber attacks and abnormal operation of switches. One operational issue that affects the topology of distribution networks is a line outage. Line outages are open lines in the distribution network that happen due to protective devices isolating some areas of the network. The cause for such isolation could be faults, anomalies, or physical topology attacks. A physical topology attack changes the dynamics of the network by physically removing bus interconnections\cite{topologyattackKar1}. Due to such isolation, the part of the network disconnected form the grid observes a loss of power. Hence, detection of such outages (outage detection) or equivalently identification of the operational topology (topology identification) is necessary to return the system to a stable operating point.

Previous studies investigated outage detection for the power transmission networks \cite{tate_overbye_1,tate_overbye_2,giannakis_line_outage,quickest_outage_detection}. However, due to the following differences between transmission and distribution networks, these methods are not best suitable to distribution networks. Firstly, transmission networks are loopy in nature while distribution networks are normally operated as radial networks. As a consequence of radial nature, power flows are unidirectional in distribution networks from  the upstream substation (root node) to downstream (load points). Secondly, the line reactance-resistance ratio in transmission networks  is  larger than that of distribution networks. This makes the standard transmission network circuit models such as the DC power flow model inadequate for distribution network studies\cite{hertem}. Hence, assuming the widespread adoption of advanced sensors such as those mentioned previously, new outage detection and topology identification methods for distribution networks are being proposed. Using time series measurements from PMUs, the authors of \cite{time_series_dist} propose a topology identification method. Network topology is identified using smart-meter data in \cite{watson}. Mixed integer quadratic programming is used for topology identification in \cite{tian} while a topology identification algorithm using voltage correlation data is proposed in \cite{backhaus}. More related to the work presented in this paper is the work in \cite{zhao,sevlian}. Utilizing power flow measurements from sensors along with load forecast statistics, the authors of \cite{zhao,sevlian} propose maximum a-posteriori (MAP) and maximum likelihood (ML) outage detection algorithms respectively to detect line outages. 

Primarily, topology identification and outage detection are performed in two stages. First is the planning stage in which sensors are placed in the distribution network with the objective of obtaining measurements that enable observability and identifiability. Second is the operation stage in which utilizing the sensor measurements, operators detect line outages using algorithms such as those mentioned above. In this study, we focus on the planning stage. We propose a cost optimal sensor placement solution for topology identification in distribution networks. Multiple sensor placement techniques exist in literature that were proposed for performing different power network tasks such as optimal sensor placement for unknown parameter estimation\cite{CISSsihag2018sequential} and false data detection\cite{CISSchattopadhyay2018attack}. Specifically for outage detection, the authors of \cite{zhao,sevlian} propose sensor placement strategies that are tailored to their corresponding outage detection methods. However these sensor placement methods have some drawbacks: dependency on load forecast statistics, neglecting zero-injection nodes in the sensor placement procedure, and being computationally-expensive due to  enumeration of all line outage scenarios. 

In this paper, we propose a cost optimal sensor placement method for outage detection in smart power distribution networks. The proposed solution is independent of load forecast statistics, considers various types of sensors and is suitable for usage with outage detection methods such as \cite{zhao,sevlian} that utilize multiple hypothesis testing to identify outages in distribution systems. We formulate the sensor placement problem as a cost optimization problem with binary decision variables. Various types of commercial sensors for distribution system monitoring applications exist in the market. Classifying these sensors into two categories: line sensors and node sensors, we consider placement of these sensors at minimal cost. Also, we take into account any zero-injection nodes present in the distribution network and ensure that our placement solution enables the detection of all outage scenarios that involve zero-injection nodes. 

The rest of this study is organized as follows. Section \ref{section2} presents the system model. Section \ref{section3} is discusses the problem formulation of our optimal sensor placement. Section \ref{section4} presents numerical results, followed by conclusions in Section \ref{section5}. 

%

\section{System Model}\label{section2}

We consider power distribution networks that have a radial (tree) topology, i.e., the power is supplied from the root of tree to the downstream branches and load points. 

\subsection{Topology of the Distribution Network}

We model the nominal (outage free) topology of the radial distribution network as a tree graph $\Gcal = \left\{V,E\right\}$ with $N$ nodes, where $V$ is the set of nodes (buses in the power  distribution system) and $E$ is the set of edges (lines in the power distribution system) of the graph. Hence, $V = \left\{1,2,\cdots,N\right\}$.  We consider node $1$ as the point of common coupling (PCC), which is the substation or the bus that connects the power distribution system under analysis to the transmission network (upstream grid). Therefore, node $1$ is the root node of the tree $\Gcal$. {Regarding the root node we make the following assumptions.

\begin{assumption} \label{ass1}
We assume that the root node is the only power source in the distribution network.
\end{assumption}

\begin{assumption} \label{ass2}
The edge that connects the root node (PCC) to the main grid carries the power supply required for the entire distribution network and is therefore most likely to experience an outage. Hence, we assume that the operator directly monitors this line and therefore we do not consider sensor placement for this edge.
\end{assumption}
}

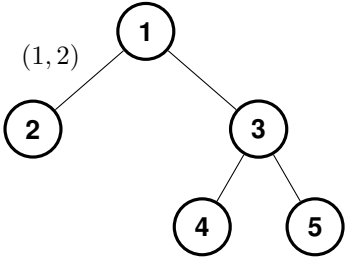
\begin{figure}[h]
	\centering
	\begin{tikzpicture}[-,level/.style={sibling distance = 3cm/#1,
		level distance =1.3cm},arn_w/.style = {treenode, circle, black, font=\sffamily\bfseries, draw=black, text width=2.0em, very thick}] 
	\node [arn_w] {1}
	child{ node [arn_w] {2}
	   edge from parent node[above left]{$(1,2)$}}
  child{ node [arn_w] {3}
	  child{ node [arn_w] {4}}
		child{ node [arn_w] {5} }};
	\end{tikzpicture}
	\caption{A distribution network represented as a tree.}
	\label{example}
\end{figure}

An edge or line between nodes $i$ and $j$ with $i$ as the starting node (i.e., power flows from bus $i$ to bus $j$) is represented as $\left(i,j\right)$. Hence, $E$ denotes set of all power distribution edges $(i,j)$ of the network. For every node $i$, $d_i$ represents the degree of the node, $C_i$ is the set of all it's children and $p_i$ is it's parent node. For every node $i$, we call the edge that connects it to it's parent node as the parent edge of node $i$ and similarly we call the edges of node $i$ that connect it to it's children as the child edges of node $i$. Also, if an edge $(i,j)$ is on the path that connects an edge $(k,l)$ to the root node $1$ then edge $(i,j)$ is said to be upstream of edge $(k,l)$ and edge $(k,l)$ is said to be downstream of edge $(i,j)$. Fig. \ref{example} illustrates the graph representation of a distribution network with $N=5$ buses. Node $1$ is the PCC or root node. The set of all nodes is defined as $V = \left\{1,2,\cdots,5\right\}$. The edge between nodes $1$ and $2$ is represented by $(1,2)$. Similarly, we can represent all the other edges in the network. Regarding the definition of each node's degree, the degrees of nodes 3 and 4 are $d_3 = 3$ and $d_4 = 1$. The set of children for node $3$ is $C_3 = \left\{4,5\right\}$ and it's parent node is $p_3 = 1$. For node $3$, the edge $(1,3)$ is it's parent edge, and the edges $(3,4),(3,5)$ are it's child edges. In order to illustrate the upstream and downstream notations, we provide the examples of line $(1,3)$ that is upstream of line $(3,4)$; and  line $(3,5)$ is downstream of line $(1,3)$. The root node $1$ has no parent node, and nodes with degree $1$ have no child nodes, i.e., $C_i = \Phi$ where $\Phi$ represents an empty set. 

\subsection{Load Model} \label{load}

Let $l_i$ denote the load consumption at node $i$ of the network. We denote the forecast of each load as $\lhat_i$ with an error $e_i = l_i-\lhat_i$. In this study, errors are assumed to be  mutually independent random variables with the zero-mean normal distribution, i.e.,   $e_i \sim N(0,\sigma_i^2)$.  The real value of load is  unknown and  considering the available load forecasts, it can be modeled as a random variable distributed as $l_i \sim N(\lhat_i,\sigma_i^2)$.Since we have load forecasts for each node, we have the vector representation

\begin{equation}
\bm{\lhat} \sim N(\bm{l},\bm{\Sigma}),
\label{noise}
\end{equation}

\noindent where $\bm{\lhat}$ and $\bm{l}$ represent the vectors of load forecasts and  true loads at each node respectively. The diagonal covariance matrix is denoted by $\bm{\Sigma}$. In most power distribution systems, there exist some nodes with zero load consumption. In this study, we refer to these nodes as zero-injection nodes. We represent the set of all zero-injection nodes in a network as $Z$. For any zero-injection node $i \in Z$, we assume $l_i = \lhat_i = \sigma_i = 0$. Regarding the set $Z$ we make the following assumption.

\begin{assumption} \label{ass3}
We assume that the set of zero-injection nodes $Z$ remains constant for all time, i.e., a zero-injection node $i \in Z$ will always have $\lhat_i = 0$ and $\sigma_i = 0$ and a nonzero-injection node $j \in V \setminus Z$ will always have $\lhat_j > 0$ and $\sigma_j \geq 0$.
\end{assumption}

It is important to note here that we do not utilize load statistics of nonzero-injection nodes during sensor placement. Hence, they will not affect the optimality of our sensor placement. However, we consider that knowledge of current load statistics is available for outage detection.

\subsection{Sensor Types}
We consider placing two categories of sensors in a distribution network. The various commercially available sensors can be classified into one of the two categories depending on their monitoring capabilities. The two categories of sensors are:

\emph{1- Line Sensor:} A line sensor is installed on an edge of a distribution tree network. In this paper we consider that the point of installation of a line sensor on an edge $(i,j)$ is towards the end of the edge connected to node $j$, and that the line sensor measures the power flow on the edge and the voltage magnitude at node $j$. 

\emph{2- Node Sensor:}  A node sensor is installed at a node of a distribution network. A micro-PMU is an example of a node sensor. In this paper, we assume that a node sensor installed at a node measures the power flow on all the edges connected to that node, as well as the voltage magnitude of that node. 

\begin{figure}[h]
	\centering
	\begin{tikzpicture}[-,level/.style={sibling distance = 3cm/#1,
		level distance =1.3cm},arn_r/.style = {treenode, circle, black, font=\sffamily\bfseries, draw=red, very thick,text width=2.0em},arn_w/.style = {treenode, circle, black, font=\sffamily\bfseries, draw=black, text width=2.0em, very thick},edg_b/.style = {edge from parent/.style = {black,thick,draw}},
		edg_r/.style = {edge from parent/.style = {green,very thick,draw}}] 
	\node [arn_r] {1}
	child{node [arn_w] {2}
	   edge from parent node[above left]{$(1,2)$}}
  child{node [arn_w] {3}
	  child{ node [arn_w] {4}}
		child[edg_r]{node [arn_w] {5}}};
	\end{tikzpicture}
	\caption{Sensor placement}
	\label{sensor}
\end{figure}
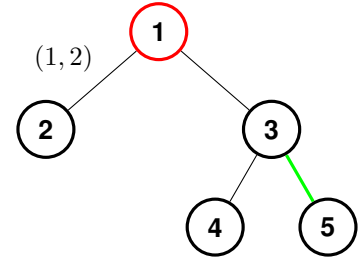

Fig. \ref{sensor} illustrates the network of Fig. \ref{example} with a line sensor (in green) on edge $(3,5)$ and a node sensor (red circle) at node $1$. We can represent the sensor placement by $\Pcal = (V_{\Pcal},E_{\Pcal})$ where the set of nodes endowed with a node sensor is $V_{\Pcal} \subseteq V$ and the set of edges endowed with a line sensor is $E_{\Pcal} \subseteq E$. For Fig. \ref{sensor}, we have $V_{\Pcal} = \left\{1\right\}$, $E_{\Pcal} = \left\{(3,5)\right\}$.

\subsection{Power Flow Model and Sensor Measurements}

In this work, we consider the linearized DistFlow equations for the power flow model \cite{dobbe}. We assume that the sensors measure only real power flows and therefore from now on we shall simply refer to real power flows as power flows. Under assumption \ref{ass1}, according to the linearized DistFlow equations we can write the true power flow $\hat{s}_{(i,j)}$ on an edge $(i,j)$ as the sum of all loads downstream of that edge, i.e., $\hat{s}_{(i,j)} = \sum_{k \in T_j}l_k$ where $T_j$ is set of all nodes in the sub-tree rooted at node $j$. 

Coming to the sensor measurements, let $\Scal_{\Pcal}$ be the set of all edges whose power flow is measured either by a node or a line sensor or both under the placement $\Pcal$. Let $\Mcal_{\Pcal}$ be the set of all nodes which have their voltage magnitude measured either by a node or a line sensor or both. The power flow measurement on an edge $(i,j) \in \Scal_{\Pcal}$ is represented as $s_{(i,j)}$. We have

\begin{equation}\label{e1}
    s_{(i,j)} = \hat{s}_{(i,j)} + n_{(i,j)} \ \forall (i,j) \in \Scal_{\Pcal}
\end{equation}
where $n_{(i,j)}$ is the sensor noise. Since $l_k = \lhat_k + e_k$, we can re-write (\ref{e1}) as

\begin{equation}\label{e2}
    s_{(i,j)} = \sum_{k \in T_j} \lhat_k + \sum_{k \in T_j} e_k + n_{(i,j)} \ \forall (i,j) \in \Scal_{\Pcal}.
\end{equation}

Similarly, the voltage magnitude measurement at a node $j \in \Mcal_{\Pcal}$ is represented as $v_j$. We have
\begin{equation}\label{e3}
    v_j = \hat{v}_j + n_j \ \forall j \in \Mcal_{\Pcal}
\end{equation}
where $\hat{v}_{j}$ is the true voltage magnitude at node $j$ and $n_j$ is the sensor noise. 

\subsection{Outage Hypotheses}

We model each line outage as a disconnected power distribution line. Due to this disconnection, the power flow on this line is zero. Each line outage disconnects the distribution network nodes that are downstream  of the outage. In other words, it breaks the distribution network into two trees: an energized tree that is supplied power through the root node; and a  disconnected tree that has no power supply. Hence, each  line outage leads to loss of power to all downstream nodes. An outage hypothesis $H$ is defined as the set of all lines in outage, i.e., $H = \left\{(i,j) \in E| \ \text{Edge} \ (i,j) \ \text{is in outage}\right\}$. Let $\Hcal$ represent set of all line outage hypotheses of the network $\Gcal$. Since there are $N-1$ lines in $\Gcal$ and simultaneous line outages are possible, the total number of possible line outage hypotheses is $|\Hcal| = 2^{N-1}$, i.e., $\Hcal =\left\{H_1,H_2,\cdots,H_{2^{N-1}}\right\}$. For example consider Fig. \ref{example} with the outage hypothesis $H_1 = \left\{(1,3)\right\}$. Under $H_1$, due to the line outage of edge $(1,3)$ the network is divided into two trees: the remaining energized tree consisting of nodes 1 and 2 connected by the edge $(1,2)$, and a disconnected un-energized tree consisting of nodes 3,4 and 5 with edges $(3,4)$ and $(3,5)$. 

Unfortunately, unique detection of every outage hypothesis $H \in \Hcal$ is not always possible. This is because some outage hypotheses might result in the same set of sensor measurements which makes them indistinguishable. For example, in Fig. \ref{sensor}, outage hypotheses $H_1 = \left\{(1,3)\right\}$ and $H_2 = \left\{(1,3),(3,4),(3,5)\right\}$ would both result in same set of sensor measurements. This is because the line outages $(3,4)$ and $(3,5)$ are downstream of line outage $(1,3)$. However, since all edges that are downstream of other outages will be disconnected from the energized tree irrespective of whether they themselves are in outage or not, and all measurements will be exactly the same either way, outage detection is restricted to a set $\Hcal_U \subseteq \Hcal$ \cite{sevlian}. $\Hcal_U$ is the set of all uniquely identifiable outages, i.e., $\Hcal_U = \left\{H \in \Hcal | \ \text{No edge in outage is downstream of another}\right\}$.

\section{Optimal Sensor Placement}\label{section3}

We formulate the problem of sensor placement for outage detection as a cost optimization problem with binary decision variables. To this end, we first provide a list of requirements that we desire our sensor placement solution to satisfy. Then, we formulate the optimization problem and illustrate how our objective function and constraints meet these requirements. The objectives of our proposed sensor placement solution are:

\begin{enumerate}
	\item Must enable real time outage detection.
	\item Must ensure cost optimality by taking into account the costs of different sensors and their installation.
	\item Must be independent of load forecast statistics so that any changes in load demand will not affect the optimality of our sensor placement.
	\item Must ensure that all outage scenarios involving edges of zero-injection nodes can be  detected. 
\end{enumerate}

To meet the above objectives, we model the sensor placement as a cost optimization problem.
Let $x_i$ represent a binary variable that corresponds to whether we place a node sensor at node $i \in V$ , i.e., $x_i = 1$, or not, i.e., $x_i = 0$. Let $\bm{x}$ represent the vector of all $x_i$. Let the binary variable $y_{(i,j)}$ represent the installation of a line sensor on line $(i,j) \in E$, i.e., $y_{(i,j)}=1$ if we install a sensor on line $(i,j)$,  or   $y_{(i,j)}=0$ if we do not. We have $\bm{y}$ as the vector all $y_{(i,j)}$. Then, the sensor placement problem is formulated as the following cost minimization problem $OP$:



\begin{align}
OP: \underset{\bm{x},\bm{y}}{\text{minimize:}}& \sum\limits_{i \in V} a_i x_i + \sum\limits_{(i,j) \in E} b_{(i,j)} y_{(i,j)} \label{obj} \\
\text{subject to:}
& \, d_1 x_1 + \! \sum\limits_{j \in C_1} x_j + \! \sum\limits_{(1,j) \in E} y_{(1,j)} \geq d_1-1 \label{c1}\\
& d_k x_k + \! \sum\limits_{j \in C_k} x_j + \! \sum\limits_{(k,j) \in E} y_{(k,j)} \nonumber \\
						 &\geq d_k-2 \ \forall k \in V-\left\{1\right\} \ \text{having} \ d_k \geq 3 \label{c2} \\
& x_k + y_{(p_k,k)} \geq 1 \ \forall k \in Z \label{c3}. 
\end{align}
where $a_i$ is the cost of installing a node sensor at node $i \in V$ and $b_{(i,j)}$ represents the cost of installing a line sensor on line $(i,j) \in E$.

Under noise free conditions, the constraints of optimization problem $OP$ guarantee that sufficient sensor data is provided so as to detect all single edge outages. Since, we restrict outage detection to $\Hcal_U$, this results in detection of all outage hypotheses $H \in \Hcal_U$ for a given radial distribution system. Since the constraints of $OP$ guarantee outage detection with absolute certainty under noise free conditions, they can also be used for outage detection under noisy conditions with an appropriate detection algorithm. However, the performance of the detection algorithm under noisy conditions will depend on the noise statistics. Now, we shall see how the constraints of $OP$ guarantee detection under noise free conditions.

Constraint (\ref{c1}) ensures that single edge outages of all child edges of the root node are all detectable. It requires that a combination of a node sensor at the root node or line sensors on edges of the root node or node sensors at child nodes of the root node, must monitor all $d_1-1$ child edges of the root node. Root node has $d_1-1$ child edges since it has one parent edge that connects it to the main grid. Monitoring of all child edges of the root node is required since these supply power from the root node to all downstream nodes. In Fig. \ref{sensor}, child edges of the root node are monitored by a node sensor at the root node. 

Constraint (\ref{c2}) ensures that edge outages of child edges of non-root nodes with degree greater than or equal to $3$ are detectable. A non-root node $k \in E$ with $d_k \geq 3$ has one parent edge and $d_k-1$ child edges. Since the parent edge is a child edge of another node, we can assume that the outages of the parent edge can be identified by a sensor upstream. Hence, we only need to ensure the identifiability of outages of the $d_k-1$ child edges. For this, it is sufficient for us to monitor $d_k-2$ child edges by sensors. Hence for a non-root node $k \in E$ with $d_k \geq 3$, constraint (\ref{c2}) requires that a combination of a node sensor at the non-root node or line sensors on edges of the non-root node or node sensors at child nodes of the non-root node, must monitor at least $d_k-2$ child edges of the non-root node. It is important to note here that since degree $2$ nodes have only one child edge, constraints (\ref{c1}) and (\ref{c2}) together inherently guarantee the outage identifiability of child edges of non-root nodes with degree $2$. 

An important requirement for our sensor placement was to ensure that outages of all edges of zero-injection nodes are identifiable. For this, in addition to constraints (\ref{c1}) and (\ref{c2}), we need constraint (\ref{c3}). Consider for example that node $3$ in Fig.\ref{sensor} is a zero-injection node and that under an outage hypothesis edge $(1,3)$ was in outage. Then the node sensor at 1 would measure zero flow on edge $(1,3)$ and a voltage at node 1, while the line sensor on edge $(3,5)$ would measure zero power flow and zero voltage. Now consider another outage hypothesis in which all both the child edges of node 3 are in outage. Even in this case, the node sensor at 1 would measure zero flow on edge $(1,3)$ and the line sensor on edge $(3,5)$ would measure zero power flow and zero voltage. Hence, the sensor placement of Fig. \ref{sensor} cannot distinguish between the two different outage scenarios. One way to distinguish the two outage scenarios is by either having a line sensor on edge $(1,3)$ or a node sensor at 3. Constraint (\ref{c3}) ensures this. 

The objective function (\ref{obj}) considers cost minimization and thereby satisfies our requirement of having a cost effective sensor placement. The costs $a_i \forall i \in V$ and $b_{(i,j)} \forall (i,j) \in E$ are user defined and can vary significantly depending on the sensor products and utility practices. Another requirement was the independence of sensor placement on load forecast statistics which $OP$ satisfies as it does not consider load forecast statistics. As mentioned before, our optimal sensor placement solution is suitable for statistical outage detection algorithms proposed in \cite{zhao,sevlian}.

\section{Simulation Results} \label{section4} 

In this section, we illustrate optimal sensor placement for a radial test feeder model with $N = 30$ nodes, with node $1$ as the root node. We find the optimal sensor placement for this network under three different  cases: Case 1, Case 2 and Case 3. Following are the details for the three cases.

\begin{enumerate}
\item In Case 1, the network has no zero-injection nodes. The cost of placing a node sensor at every node is $2$ and the cost of placing a line sensor on every edge is $1$, i.e., $a_i = 2 \forall i \in V$ and $b_{(i,j)} = 1 \ \forall (i,j) \in E$.
\item In Case 2, the network again has no zero-injection nodes but we now have $a_i = 3 \forall i \in V$ and $b_{(i,j)} = 1 \ \forall (i,j) \in E$.
\item In Case 3, the network has zero injection nodes, $Z = \left\{3,4,11\right\}$. As in Case 1, we have $a_i = 2 \forall i \in V$ and $b_{(i,j)} = 1 \ \forall (i,j) \in E$. 
\end{enumerate}

\begin{figure}
\begin{center}
\includegraphics[width = 0.5\textwidth, height = 0.4\textwidth]{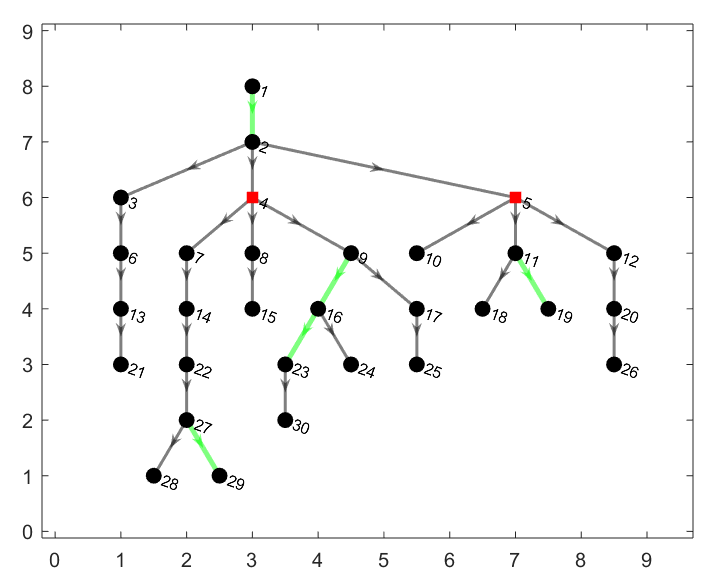}
\caption{Test feeder: Case 1}\label{s1}
\end{center}
\end{figure}

\begin{figure}
\begin{center}
\includegraphics[width = 0.5\textwidth, height = 0.4\textwidth]{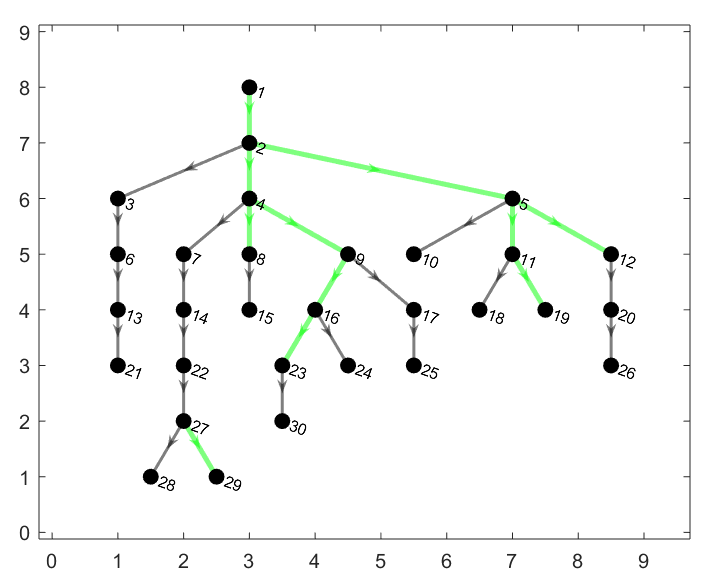}
\caption{Test feeder: Case 2}\label{s2}
\end{center}
\end{figure}

\begin{figure}
\begin{center}
\includegraphics[width = 0.5\textwidth, height = 0.4\textwidth]{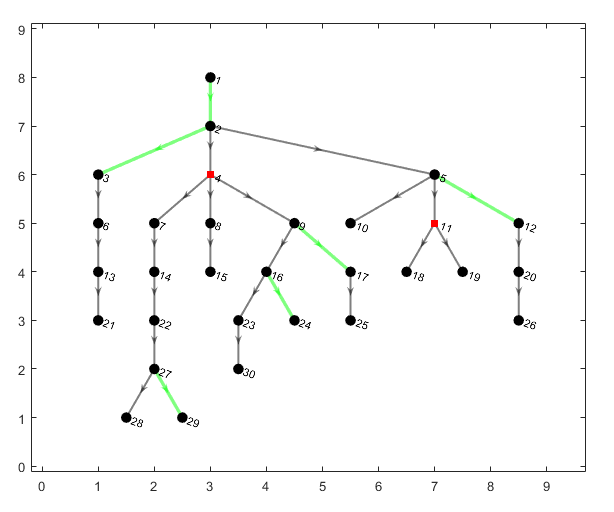}
\caption{Test feeder: Case 3}\label{s3}
\end{center}
\end{figure}

All results presented here were computed using MATLAB on a 3.4GHz Intel(R) Core(TM) i7-2600 processor with 8 GB of RAM. In the illustrated results, an edge in green represents a line flow sensor and a node in red represents a node sensor. The optimal sensor placement for Case 1, Case 2 and Case 3 are illustrated in Fig. \ref{s1}, Fig. \ref{s2} and Fig. \ref{s3} respectively. Table \ref{ts1} compares the three cases in terms of the number of node and line sensors. Comparing Case 1 and Case 2, it can be concluded that the sensor placement solution is different for the two cases because of the difference in sensor costs. The placement solution does not place any node sensors in Case 2. This is due to the fact that the relative cost of a node sensor to a line sensor is higher in Case 2 as compared to Case 1. Similarly, the sensor placement solution for Case 3 is different from Case 1 even though in both cases the costs are the same. This is because of the presence of zero-injection nodes in Case 3. The placement solution has to satisfy (\ref{c3}) for Case 3 and hence results in a different optimum sensor placement.

\begin{table}
 \caption{Sensor placement results for test feeder with 30 nodes}
\label{ts1}
\begin{center}
 \begin{tabular}{| c |c | c | c |} 
\hline
Case & Number of & Number of \\
 & Node Sensors & Line Sensors \\
\hline
Case 1 & 2 & 5 \\
\hline
Case 2 & 0 & 11 \\
\hline
Case 3 & 2 & 6 \\
\hline
\end{tabular}
\end{center}
\end{table}

\section{Conclusions}\label{section5}

In this paper, we have proposed a novel formulation of a cost optimal sensor placement for outage detection in power distribution systems. We have formulated the sensor placement problem as a cost minimization problem subject to constraints that enable detectability of all line outages. Our model covers all nodes in the network, including  zero-injection nodes. We presented numerical results that illustrate the proposed placement solution for a test feeder and which also highlight some important characteristics of our sensor placement solution. Currently, we are investigating algorithms that can efficiently solve the formulated  sensor placement optimization problem.

\bibliographystyle{IEEEtran}
\bibliography{References1}
\vfill

\end{document}